\documentclass{iau}

\usepackage{amsmath}
\usepackage{graphicx}
\usepackage{multirow}

\begin{document}

\lefttitle{Caroline Mannes}
\righttitle{The impact of stellar mergers on  Type~IIP supernovae}
\title{Binary-induced mass increase and helium enrichment of the envelopes of Type~IIP supernova progenitors}
\author{Caroline Mannes$^{1,2}$, Norbert Langer$^{1,2}$, Andrea Ercolino$^{1}$}
\affiliation{email: \email{cmannes@astro.uni-bonn.de}\\ $^1$Argelander-Institut für Astronomie, Universität Bonn,
Auf dem Hügel 71, 53121 Bonn, Germany}
\affiliation{$^2$Max-Planck-Institut für Radioastronomie,
Auf dem Hügel 69, 53121 Bonn, Germany}

\jnlPage{1}{7}
\jnlDoiYr{2021}
\doival{10.1017/xxxxx}

\aopheadtitle{Proceedings IAU Symposium}
\editors{editors}

\begin{abstract}
Binary interactions are expected to play a major role in shaping the progenitor population of core-collapse supernovae, particularly in the era of deep, high-cadence transient surveys. Using \texttt{ComBinE}, a rapid binary population synthesis code combining tabulated stellar structure models with physically motivated mass-transfer stability criteria, we find that $\sim40\%$ of Type~II supernova progenitors at Milky Way metallicity originate from mergers between a post-main-sequence star and a main-sequence companion following unstable mass transfer. Compared to single-star progenitors with the same helium-core mass, these merger products have more massive hydrogen-rich envelopes and broader envelope-mass distributions. Their envelopes are helium-enriched, with relative enhancements of up to $80\%$ and helium mass fractions reaching 0.4. These differences are expected to affect Type~IIP light curves and should be considered in future progenitor and explosion models. Constraining the merger-induced Type~IIP fraction with facilities such as LSST, Roman, and the ELT will provide insights into binary evolution physics.
\end{abstract}


\maketitle

\section{Introduction}

Type~II supernovae are the most common type of core-collapse supernovae \citep{Pessi2025}. Their spectra are characterized by prominent hydrogen lines. The class of Type~IIP supernovae exhibits an extended plateau in the optical light curve, caused by hydrogen recombination in the expanding ejecta \citep{Popov1993}. Because of their high occurrence rate and the connection between the explosion and light-curve properties and the structure of the progenitor star, Type~IIP supernovae provide an important opportunity to study the late stages of massive stellar evolution. Massive stars rarely evolve in isolation. Observations indicate that roughly 70\% of O-type stars experience binary interaction during their lifetime \citep{Sana2012}. Binary evolution provides a prominent channel for producing stripped-envelope supernovae, whose progenitors have lost most or all of their hydrogen-rich envelope prior to core collapse \citep{Smith2011}. 

The influence of binary interactions on hydrogen-rich supernovae has received less attention. Population synthesis predictions find that single stars only make up less than half of the progenitors of Type~II supernovae, while the rest are made up of accretors and merger products in binaries \citep{Ercolino2026}. Recent comparisons between observed and simulated Type~IIP light curves have revealed tensions that suggest current progenitor models may not fully capture the diversity of hydrogen-rich supernovae. Since most progenitor models employed in these studies are based on single-star evolution, an important source of diversity may be missing \citep{Fang2025}. 

In particular, stellar mergers can substantially alter the hydrogen-rich envelope while leaving the helium core unchanged, producing progenitors with structural properties that are inaccessible through single-star evolution \citep{Schneider2025}. Such changes are expected to influence the resulting light curve. In this work, we investigate the impact of post-main-sequence stellar mergers on the progenitors of Type~II supernovae using rapid binary population synthesis calculations. We estimate the fraction of Type~II supernovae that originate from merger products and examine how mergers modify the masses and helium mass fractions of hydrogen-rich envelopes of progenitor stars.

\section{Methods}
We performed binary population synthesis calculations using \texttt{ComBinE} \citep{Kruckow2018}, a rapid binary evolution code based on \cite{TaurisBailes1996} and \cite{VossTauris2003}. \texttt{ComBinE} employs updated prescriptions for Roche-lobe overflow and mass-transfer stability based on \cite{schurmannMTa} and \cite{schurmann2025}. The code adopts physically motivated criteria for unstable mass transfer and mergers. Systems merge either when thermal-timescale mass transfer causes the accretor to expand into contact and overflow through the outer Lagrange point, or when Roche-lobe overflow from a donor with a deep convective envelope becomes dynamically unstable. Mass transfer is only partially conservative, with the accretion efficiency depending on the accretor mass \citep{schurmann2025}. This influences the stability of mass transfer given the response of the companion's radius to the accretion \citep{schurmannMTa}.
For Milky Way metallicity ($Z_\mathrm{MW}$), we simulated a population of $100,000$ binary systems with primary zero-age main-sequence (ZAMS) masses in the range $9$--$30\,M_\odot$. We investigate systems in which the primary star transfers mass to its main-sequence companion after evolving off the main sequence and subsequently undergoes unstable mass transfer, leading to a common-envelope phase. If the envelope is not successfully ejected, the binary merges and produces a single, giant star (the 'merger product'). We approximate post-main-sequence mergers by adding the entire secondary star to the hydrogen-rich envelope of the primary and assume instantaneous, homogeneous mixing of the envelope. Consequently, the helium-core mass of the merger product is equal to the helium-core mass of the primary at the time of the merger, $M_\mathrm{m,HeC} = M_\mathrm{1,HeC}$, while its envelope mass is the sum of the primary's pre-merger hydrogen-rich envelope mass and the full mass of the secondary star, $M_\mathrm{m,env} = M_\mathrm{1,env} + M_\mathrm{2,tot}$.
To quantify the impact of binary mergers on Type~II supernovae progenitors, we compare each merger product, immediately after the merger, to a single-star model with the same helium-core mass. Because the helium-core mass remains unchanged during the merger, this comparison directly quantifies the increase in the hydrogen-rich envelope mass contributed by the secondary star.
Mass loss during the merger process and subsequent evolution of the merger product towards core collapse is not considered in this work (see Sect.~\ref{sec:dis}). 
The envelope helium mass fraction is computed assuming complete mixing. Since the secondary star's central regions are enriched in helium, the merger produces envelopes that are both more massive and helium enriched compared to those of equivalent single stars.

\section{Results}
Assuming an intrinsic binary fraction of 70\%, we find that binary mergers contribute a substantial fraction of the Type~II supernovae progenitor population. For Milky Way metallicity, approximately 40\% of Type~II progenitors originate from a merger between a post-main-sequence star and a main-sequence companion following unstable mass transfer. This number is in qualitative agreement with the results from detailed models by \cite{Ercolino2026}.
 
\begin{figure}
    \centering
   \includegraphics[scale=.4]{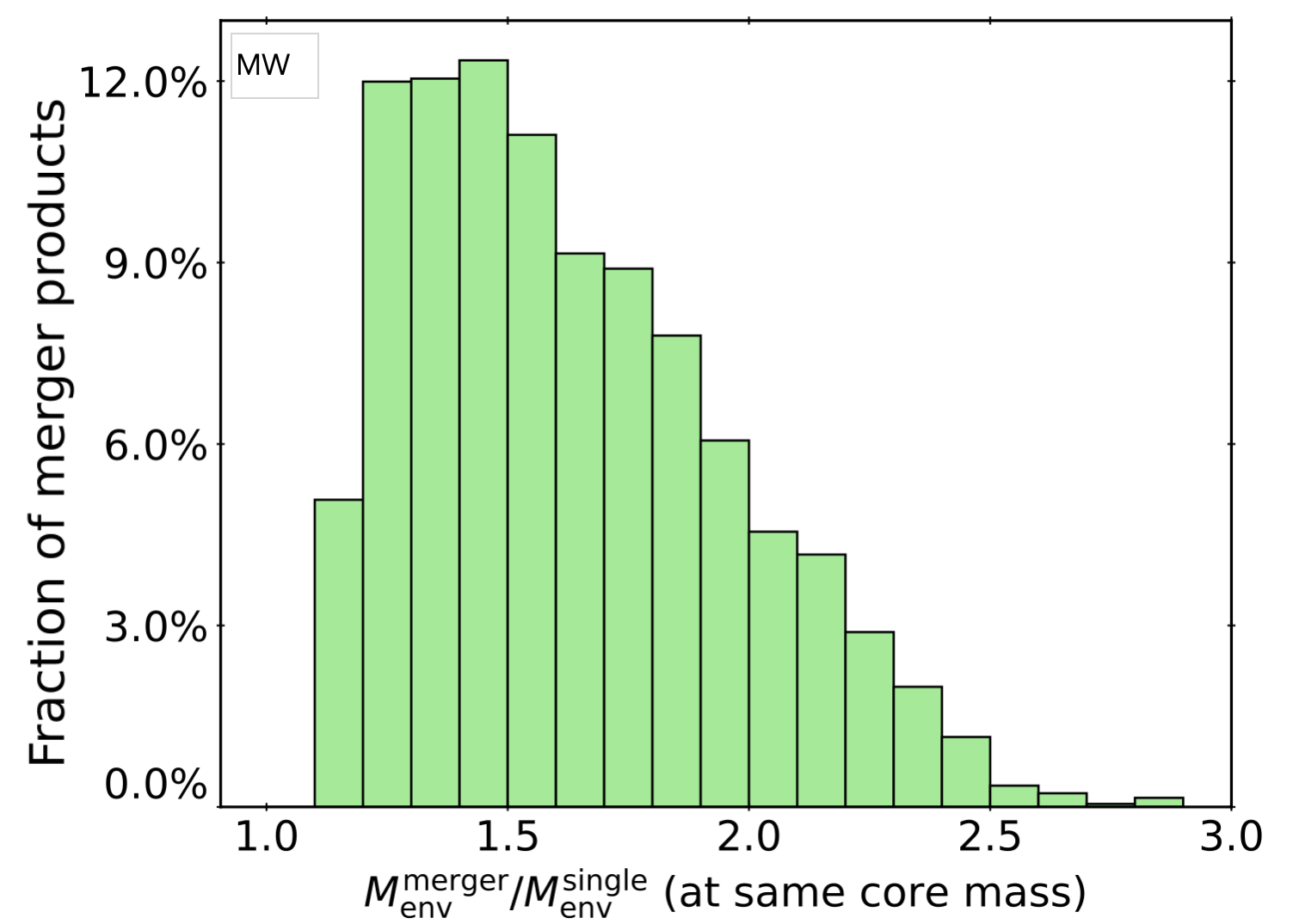}
   \caption{Distribution of the relative increase in hydrogen-rich
    envelope mass of merger products immediately after
    the merger compared to single-star progenitors
    of the same helium-core mass, at Milky Way metallicity. On average, envelope masses increase by $\sim\!64\%$ relative to single stars of the same helium-core mass, with maximum enhancements reaching $\sim\!180\%$. Most mergers, occurring at low mass ratios, produce only modest enhancements, while the largest increases arise from the less common mergers with mass ratios close to unity.}
   \label{fig:enhancement}
\end{figure}

Figure~\ref{fig:enhancement} shows the increase in hydrogen-rich envelope mass of merger products relative to single-star progenitors with the same helium-core mass. On average, mergers increase the envelope mass by approximately $64\%$, with maximum enhancements reaching about $180\%$. Most mergers occur at low mass ratios and therefore produce relatively modest increases, whereas the largest enhancements arise from the less common mergers with mass ratios close to unity, where the companion contributes a comparable amount of hydrogen-rich material to the envelope.

\begin{figure}
    \centering
   \includegraphics[scale=.4]{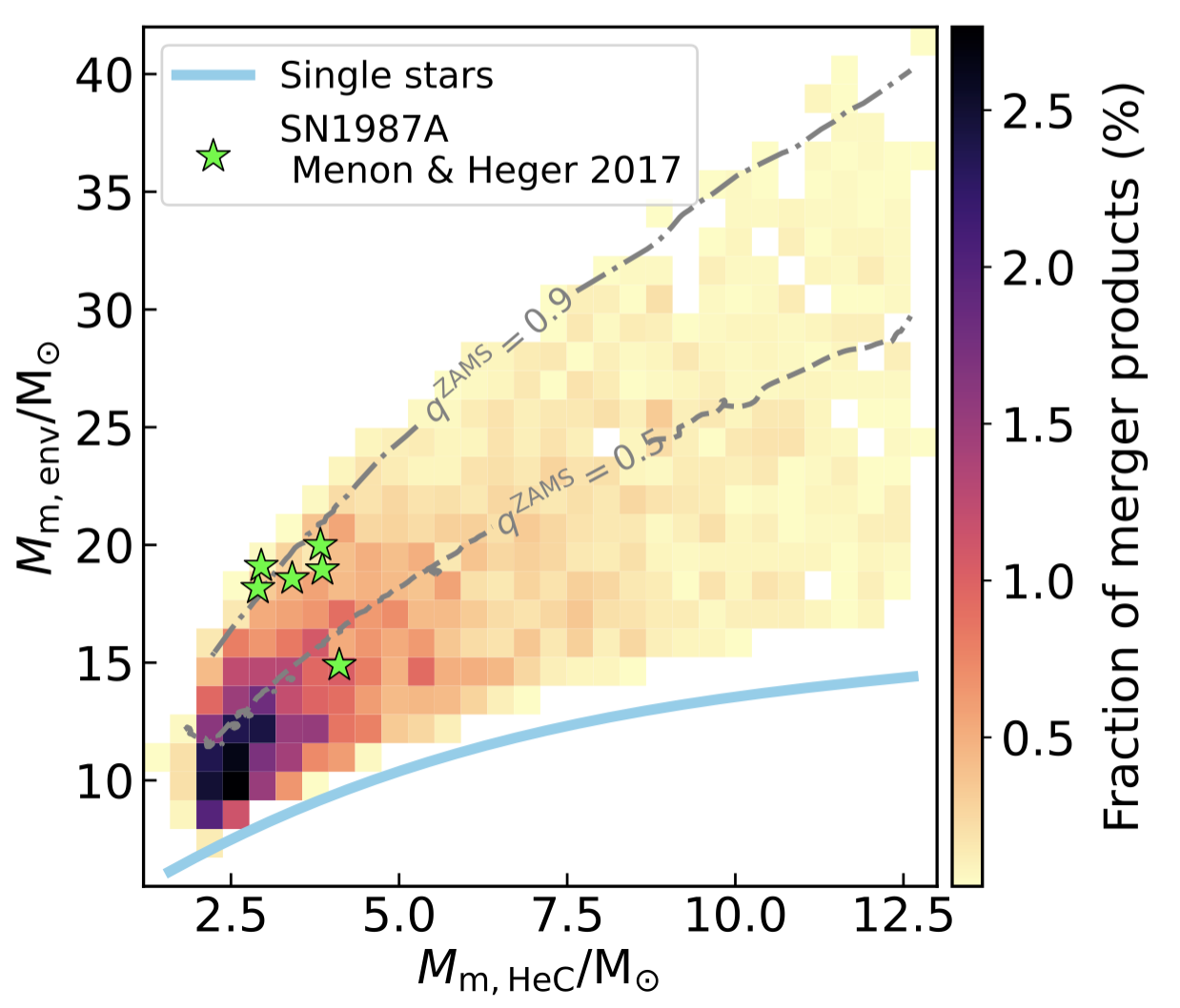}
   \caption{Hydrogen-rich envelope mass versus helium-core mass of merger products at Milky Way metallicity, color-coded by their relative frequency. The light blue line marks the single-star relation between envelope mass and helium-core mass at the same evolutionary point as the merger product, i.e. the primary without the secondary's mass added. Dashed gray lines show locations of constant ZAMS mass ratio ($q^{\rm ZAMS}=0.5$ and $0.9$), illustrating that envelope enhancement increases as the mass ratio approaches unity. Green stars indicate progenitor model parameters proposed for SN~1987A by \cite{Menon2017}. At fixed helium-core mass, merger products systematically populate envelope masses well above the single-star relation.}
   \label{fig:envmass}
\end{figure}

Figure~\ref{fig:envmass} compares the hydrogen-rich envelope masses of merger products and single-star progenitors. Unlike single-star progenitors, which have a well-defined hydrogen-rich envelope mass at a given helium-core mass, merger products exhibit a broad distribution of envelope masses that extends to significantly larger values. Binary mergers therefore introduce a larger diversity of progenitor envelope structures than expected from single-star evolution alone.

\begin{figure}
    \centering
   \includegraphics[scale=.4]{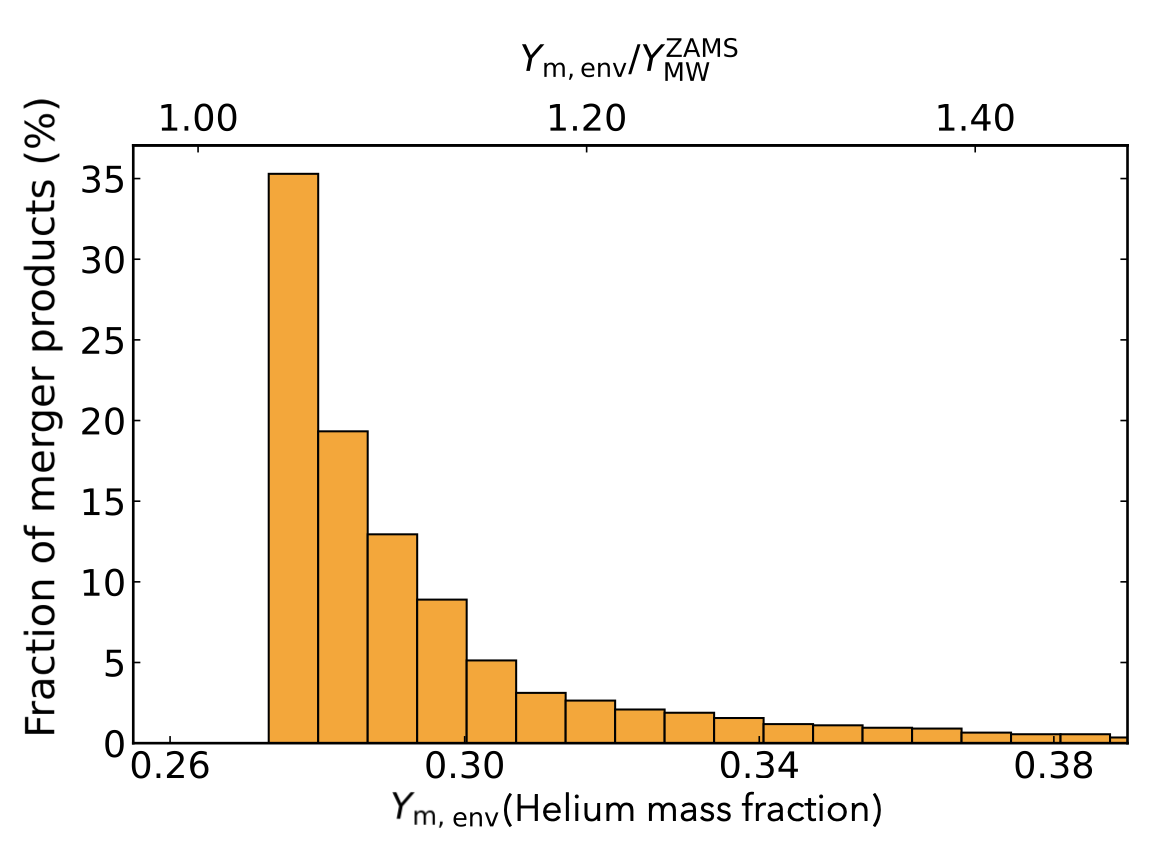}
   \caption{Distribution of helium mass fractions in the hydrogen-rich envelopes of merger products at Milky Way metallicity (bottom axis). The top axis shows the same quantity normalized to the ZAMS helium mass fraction at Milky Way metallicity, $Y_{\rm ZAMS}^{\rm MW}$, giving the enrichment factor relative to the pre-merger composition. Complete mixing of the primary envelope with the main-sequence companion produces systematically helium-enriched envelopes. The distribution peaks around moderate enrichment values, with an average enhancement of $\sim\!12\%$. A small fraction of systems ($\sim\!1\%$) extends beyond the plotted range, reaching enrichment factors of up to $\sim\!80\%$.}
   \label{fig:helium}
\end{figure}

The distribution of helium mass fractions in the hydrogen-rich envelopes of merger products is shown in Figure~\ref{fig:helium}. Compared to single-star progenitors, mergers produce systematically helium-enriched envelopes as a consequence of mixing helium-rich material from the interior of the main-sequence companion and the primary's inner regions around the core into the merger product's envelope. The average helium enrichment is approximately $12\%$. The distribution shown in Figure~\ref{fig:helium} covers the bulk of the systems, while a small fraction ($\sim\!1\%$) lies beyond the plotted range. These extreme merger products exhibit helium mass fractions of up to $\sim\!0.4$, corresponding to enrichments of up to $\sim\!80\%$ relative to the ZAMS composition. Such changes in helium mass fraction are expected to modify the opacity and recombination properties of the envelope, potentially influencing the resulting Type~IIP light curves \citep{Kasen2009}.

\section{Discussion and Conclusions}
\label{sec:dis}
Several simplifying assumptions remain in the present work. We neglect mass loss during the merger, assume instantaneous homogeneous mixing of the envelope, and do not evolve merger products beyond the point of merging, i.e. their core growth between the time of the merger and core collapse remains to be modeled. These assumptions therefore represent an upper limit to the structural changes induced by mergers.
Our results indicate that binary mergers constitute an important evolutionary channel for Type~IIP supernova progenitors. While the helium-core masses remain unchanged, the hydrogen-rich envelopes differ substantially from those of equivalent single stars, both in mass and composition. These structural differences are likely to influence the explosion and observable properties of Type~IIP supernovae, particularly during the plateau phase. Binary mergers should therefore be considered in future progenitor and light-curve calculations of Type~IIP supernovae.

\end{document}